\documentclass[graybox]{svmult}

\usepackage{mathptmx}       
\usepackage{helvet}         
\usepackage{courier}        
\usepackage{type1cm}        
\usepackage{makeidx}         
\usepackage{graphicx}        
\usepackage{multicol}        
\usepackage[bottom]{footmisc}

\usepackage{amsmath} 
\usepackage{amssymb} 

\DeclareMathOperator*{\tra}{\,^{\mbox{\tiny T}}\!}
\newcommand{\bmat}{\left[\begin{matrix}}
\newcommand{\emat}{\end{matrix}\right]}
\makeindex             

\begin{document}

\title*{Noise-induced Stop-and-Go Dynamics}
\author{Antoine Tordeux, Andreas Schadschneider and Sylvain Lassarre}
\institute{Antoine Tordeux \at Forschungszentrum J\"ulich and
  University of Wuppertal, \email{a.tordeux@fz-juelich.de}
\and Andreas Schadschneider \at University of Cologne,
\email{as@thp.uni-koeln.de}
\and Sylvain Lassarre \at IFSTTAR COSYS GRETTIA, 
\email{sylvain.lassarre@ifsttar.fr}}
%
%
\maketitle

\abstract{Stop-and-go waves are commonly observed in traffic and
  pedestrian flows.  In traffic theory they are described by phase
  transitions of metastable models.  The self-organization phenomenon
  occurs due to inertia mechanisms but requires fine tuning of
  the parameters.  Here, a novel explanation for stop-and-go waves
  based on stochastic effects is presented for pedestrian dynamics.
  We show that the introduction of specific coloured noises in a
  stable microscopic model allows to describe realistic pedestrian 
	stop-and-go behaviour without requirement of metastability and phase
  transition. We compare simulation results of the stochastic model to 
	real pedestrian trajectories and discuss plausible values for the 
	model's parameters.}

\section{Introduction}
\label{sec:1}

Stop-and-go waves in traffic flow is a fascinating collective
phenomenon that attracted the attention of scientists for several
decades \cite{Herman1959,Kerner1997,Orosz2009} (see
\cite{Chowdhury2000,Kernerbook} for reviews).  Curiously, congested flows
self-organise in waves of slow and fast traffic (stop-and-go) instead
of streaming homogeneously.  Stop-and-go dynamics are observed in road
traffic, bicycle and pedestrian movements \cite{Zhang2014} in reality
as well as during experiments, where the disturbance due to the
infrastructure cannot explain their presence \cite{Sugiyama2008}.
Besides its scientific interest, such self-organisation phenomena
impact transportation networks in terms of safety, economy, and
comfort.

Stop-and-go behaviours are often analysed with microscopic, mesoscopic
(kinetic) and macroscopic models based on non-linear differential
systems (see for instance \cite{Bando1995,Helbing1998,Colombo2003}),
but also with discrete models like cellular automata.  The models
based on systems of differential equations have homogeneous
equilibrium solutions that can be unstable for certain values of the
parameters.  Periodic or quasi-periodic solutions in unstable cases
describe non-homogeneous dynamics, with potentially stop-and-go waves
for fine tuning of the parameters, while the stable cases describe
homogeneous dynamics.
 
Phase transition and metastability in self-driven dynamical systems
far from the equilibrium 
are commonly observed in physics, theoretical biology or social science
\cite{Ben-Jacob1994,Vicsek1995,Bussemaker1997,Buhl2006,Hermann2012}. 
In traffic, typical continuous models are inertial second order
systems based on relaxation processes.  Stop-and-go and matching to
Korteweg--de Vries (KdV) and modified KdV soliton equations occur when
the inertia of the vehicles exceed critical values
\cite{Bando1995,Muramatsu1999,Tomer2000}.  Empirical evidence for
phase transitions in traffic, like hysteresis or capacity drop, have
been observed in real data as well as during experiments
\cite{Kerner1997,Sugiyama2008}.  Yet the number of phases and their
characteristics remain actively debated \cite{Treiber2010}.

Some studies describe pedestrian stop-and-go dynamics by means of, as
traffic models, instability and phase transitions
\cite{Portz2010,Moussaid2011,KUANG2012,Lemercier2012}.  However, to
our knowledge, empirical evidence for phase transitions and
metastability have not been observed for pedestrian flow.  Pedestrian
dynamics shows no pronounced inertia effect since human capacity
nearly allows any speed variation at any time.  Furthermore, pedestrian
motion does not show mechanical delays.  Nevertheless, stop-and-go
behaviour is observed at congested density levels \cite{Seyfried2010,Zhang2014}.

In this work, we propose an novel explanation of stop-and-go phenomena
in pedestrian flows as a consequence of stochastic effects.  We first
present statistical evidence for the existence of Brownian noise in
pedestrian speed time-series. 
Then a microscopic model composed of a
minimal deterministic part for the convection and a relaxation process
for the noise is proposed and analysed.  Simulation results show that
the stochastic approach allows to describe realistic pedestrian
stop-and-go dynamics without metastability and fine tuning of the
parameters.


\section{Definition of the stochastic model}
\label{sec:2}

Stochastic effects can have various roles in the dynamics of
self-driven systems \cite{Haenggi2007}.  Generally speaking, the
introduction of white noise in models tends to increase the disorder
in the system \cite{Vicsek1995} or to prevent self-organisation
\cite{Helbing2000}, while coloured noises can affect the dynamics and
generated complex patterns \cite{Arnold1978,Castro1995}. 
Coloured noise have been observed in human response \cite{Gilden1995,Zgonnikov2014}.
Pedestrian as well as driver behaviours result from complex human cognition. 
They are intrinsically stochastic in the sense that the deterministic modelling of the driving, 
i.e.\ the modelling of the human cognition composed of up to 10$^\text{11}$ neurons \cite{Williams1988}, 
is not possible. 
Furthermore the behaviour of a pedestrian, as well as a driver, may be influenced by multiple factors, e.g.\ experience, culture, 
environment, psychology, etc.   
Stochastic effects and notion of noise are the main emphasis of many pedestrian or road traffic modelling approaches 
(see, e.g., white noises \cite{Helbing1995,Tomer2000}, pink-noise \cite{Takayasu1993}, 
action-point \cite{Wagner2006}, or again inaccuracies or risk-taking behavior \cite{TREIBER2006,Hamdar2015}). 

\begin{figure}[!ht]
\sidecaption[t]
\includegraphics{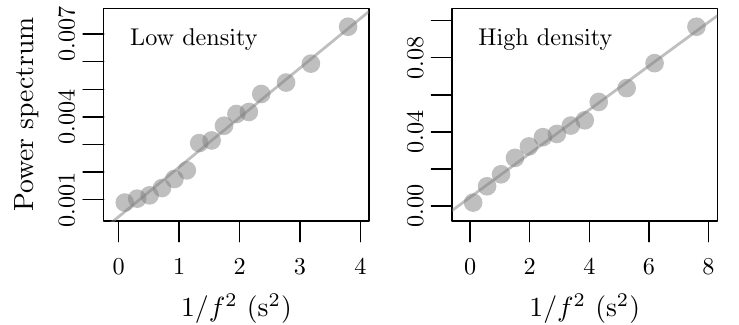}
\caption{Periodogram power spectrum estimate for the speed time-series
  of pedestrians at low and high density levels.  The power spectrum
  is roughly proportional to the inverse of square frequency $1/f^2$.
  This is a typical characteristic of a Brownian noise.}
\label{fig:0}       
\end{figure}

Fig.~\ref{fig:0} presents statistical evidence for the existence of a
Brownian noise in time series of pedestrian speeds.  Such a noise has
power spectral density (PSD) proportional to the inverse of the square
noise frequency $1/f^2$.  A characteristic linear tendency is observed
independent of the density (see \cite{Tordeux2016} for details
on the data).  Such a noise with exponentially decreasing
time-correlation function can be described by using the
Ornstein-Uhlenbeck process (see e.g.\ \cite{lindgren2013}).

We denote in the following $x_k(t)$ the curvilinear position of the
pedestrian $k$ at time $t$.  The pedestrian $k+1$ is the predecessor
of $k$.  The model is the Langevin equation
\begin{equation}
\left\{\begin{array}{lcl}
\mbox d x_k(t)&=&V\big(x_{k+1}(t)-x_k(t)\big)\,\mbox dt+\varepsilon_k(t)\,\mbox dt,\\[.5mm]
\mbox d\varepsilon_k(t)&=&-\frac1\beta\varepsilon_k(t)\,\mbox dt+\alpha\,\mbox d W_k(t).\end{array}\right.
\label{mod}
\end{equation} 
Here $V:s\mapsto V(s)$ is a differentiable and non-decreasing
\emph{optimal velocity} (OV) function for the convection \cite{Bando1995}, while
$\varepsilon_k(t)$ is a noise described by the Ornstein-Uhlenbeck
stochastic process.  In the following, an affine function
$V(s)=\frac1T(s-\ell)$ is used with $T$ the time gap between the
agents and $\ell$ their size.  The quantities $(\alpha,\beta)$ are
positive parameters related to the noise.  $\alpha$ is the volatility
while $\beta$ is the noise relaxation time.  $W_k(t)$ is a Wiener
process.  Note that alternatively, the model can be defined as the
special stochastic form of the Full Velocity Difference model
\cite{Jiang2001}
$\ddot x_k=\big[V(x_{k+1}-x_k)-\dot
x_k\big]/\beta+V'(x_{k+1}-x_k)(\dot x_{k+1}-\dot x_k)+\alpha\xi_k$,
where $\xi_k$ is a white noise.

Considering a line of $n$ agents with periodic boundary conditions,
the system is
\begin{equation}
\left\{\begin{array}{l}
\mbox d \eta(t)=(A\eta(t)+a)\,\mbox dt+b\,\mbox d w(t)\\[1.5mm]
\mbox{with$\quad\eta(t)=\tra\big(x_1(t),\varepsilon_1(t),\ldots,x_n(t),
\varepsilon_n(t)\big)\in\mathbb R^{2n}$}\\[.25mm]
\mbox{$A=$\small$\bmat R&S&\\[-2.4mm]&\ddots&\\[-2.4mm]S
&&R\emat\in\mathbb R^{2n\times2n}$\normalsize~ with~ 
$R=$\small$\bmat -1/T&1\\0&-1/\beta\emat$
\normalsize~ and~ $S=$\small$\bmat 1/T&0\\0&0\emat$}\\[2.75mm]
\mbox{$a=-\frac{\ell}T\tra(1,0,\ldots,1,0)\in\mathbb R^{2n}$ 
~and~ $b=\alpha\tra(0,1,\ldots,0,1)\in\mathbb R^{2n}$.}\end{array}\right.
\label{syst}\end{equation}
Here $w(t)$ is a $2n$-vector of independent Wiener processes.  Such a
linear stochastic process is Markovian. 
It has a normal distribution with expectation $m(t)$ and
variance/covariance matrix $C(t)$ such that, by using the Fokker-Planck
equation,
\begin{equation}
\dot m(t)=Am(t)+a\qquad\mbox{and}\qquad\dot C(t)=AC(t)+(C(t))\!\tra A
+\mbox{diag}(b)\quad
\end{equation} 
with $m(0)=\eta_0$ and $C(0)=0$.  The expected value $m(t)$ is
asymptotically the homogeneous solution for which
$x_{k+1}(t)-x_k(t)=L/n$ and $\varepsilon_k(t)=0$ for all $k$ and
$t$, 
$L$ being the length of the system.  The roots $(\lambda_1,\lambda_2)$
of the characteristic equation for the homogeneous configuration
\begin{equation}\textstyle
\big[\lambda+\frac1T\big(1-e^{i\theta}\big)\big]\big[\lambda+1/\beta\big]=0,
\end{equation}
have strictly negative real parts 
for any $\theta\in(0,2\pi)$.  Therefore the homogeneous solution is
stable for the system (\ref{syst}) for any values of the parameters.
Note that the more unstable configuration is the one with maximal
period for which $\theta\rightarrow0$.

\section{Numerical experiments}
\label{sec:3}

The system (\ref{syst}) is simulated using an explicit Euler-Maruyama
numerical scheme with time step $\delta t=0.01$~s.  The parameter
values are $T=1$~s, $\ell=0.3$~m, $\alpha=0.1$~ms$^{-3/2}$ and
$\beta=5$~s.  Such values are close to the statistical estimates for
pedestrian flow presented in \cite{Tordeux2016}.  The length of the
system is $L=25$~m, corresponding to the experimental situation, and
the boundary conditions are periodic.

\begin{figure}[!ht]
\sidecaption[t]
\includegraphics{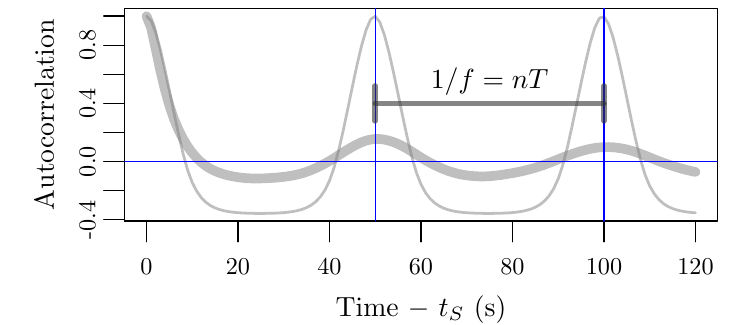}
\caption{Mean time-correlation function of the distance spacing for the
  stochastic and deterministic models in stationary state.  The same
  period $1/f=nT$ for the stop-and-go waves is observed.  The
  simulation time to consider the system stationary is $t_S=\,$2$\cdot$10$^\text5$~s. }
\label{fig:2}       
\end{figure}

Simulations are carried out for systems with $n=25$, $50$ and $75$ agents with the stochastic model Eq.~(\ref{syst}) 
and the unstable deterministic optimal velocity model introduced in \cite{Tordeux2014} from jam initial condition.
The mean time-correlation functions for the distance spacing in stationary state (i.e.\ for large simulation time) are presented in Fig.~\ref{fig:2}. 
The peaks of the time-correlations match for both stochastic and deterministic models, i.e.\ the frequency of the stop-and-go waves 
are the same. 
A wave propagate backward in the system at a speed $c=-\ell/T$ while vehicles travel in average at the 
speed $v=(L/n-\ell)/T$. 
In adequacy with the theory, the wave period for any agent is $(v-c)/L=nT$. 

\begin{figure}[!ht]
\includegraphics{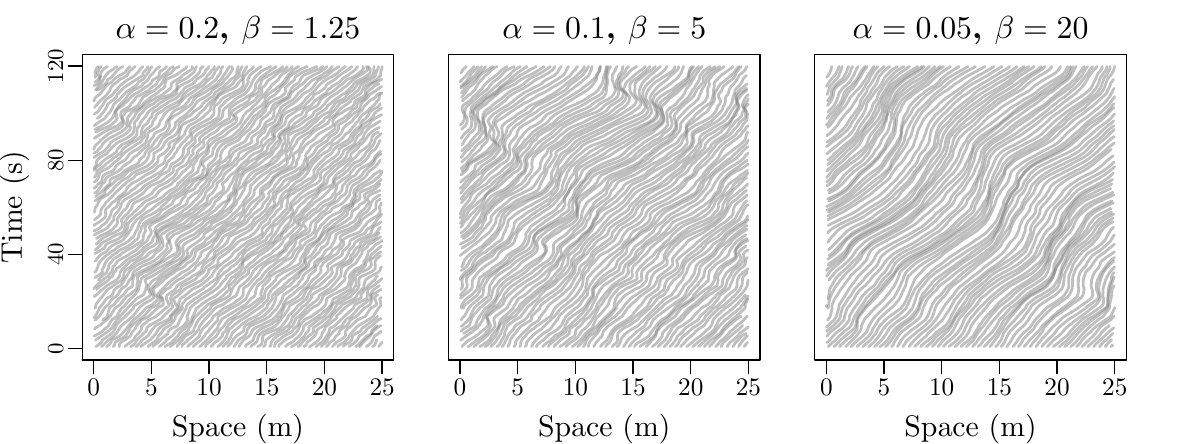}
\caption{Simulated trajectories for different values of the noise
  parameters (units: $\alpha$ in ms$^{-3/2}$, $\beta$ in s).  The
  initial configuration is homogeneous. Here $n=50$ agents are
  considered.}
\label{fig:3}       
\end{figure}

\begin{figure}[!ht]
\sidecaption[t]
\includegraphics{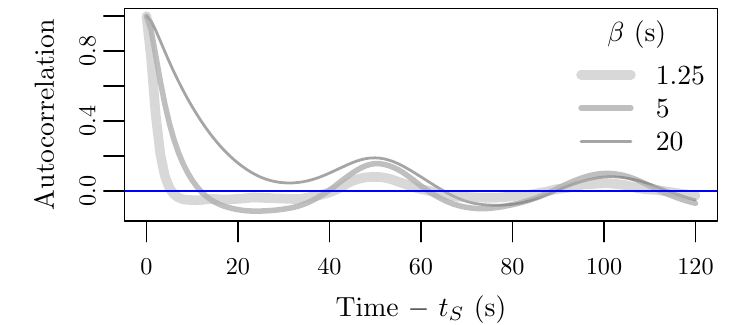}
\caption{Mean time-correlation function of the distance spacing in the
  stationary state for different values of the noise parameters.  
  The noise parameters do not influence the frequency of the
  waves (that only depends on $n$ and $T$).
  The simulation delay time is  $t_S=\,$2$\cdot$10$^\text5$~s.}
\label{fig:4}       
\end{figure}

Some experiments are carried out for different values of the noise
parameters $\alpha$ and $\beta$.  In Fig.~\ref{fig:3}, the
trajectories of $50$ agents are presented for $\alpha=0.05$, $0.1$ and
$0.2$~ms$^{-3/2}$ while $\beta=1.25$, $5$ and $20$~s (the values are
set in order to keep the amplitude of the noise 
$\sigma=\alpha\sqrt{\beta/2}$ constant).  The noise tends to be white when
$\beta$ is low, while the noise autocorrelation is high for large
relaxation times $\beta$.  Unstable waves emerge locally and disappear
when $\beta$ is small (i.e.\ for a white noise, see Fig.~\ref{fig:3},
left panel), while stable waves with large amplitude occur for high
$\beta$ (Fig.~\ref{fig:3}, right panel).  The parameters of the noise
influence the amplitude of the time-correlation function, but not the
frequency that only depends on the parameters $n$ and $T$, see
Fig.~\ref{fig:4}.

\begin{figure}[!ht]
\includegraphics{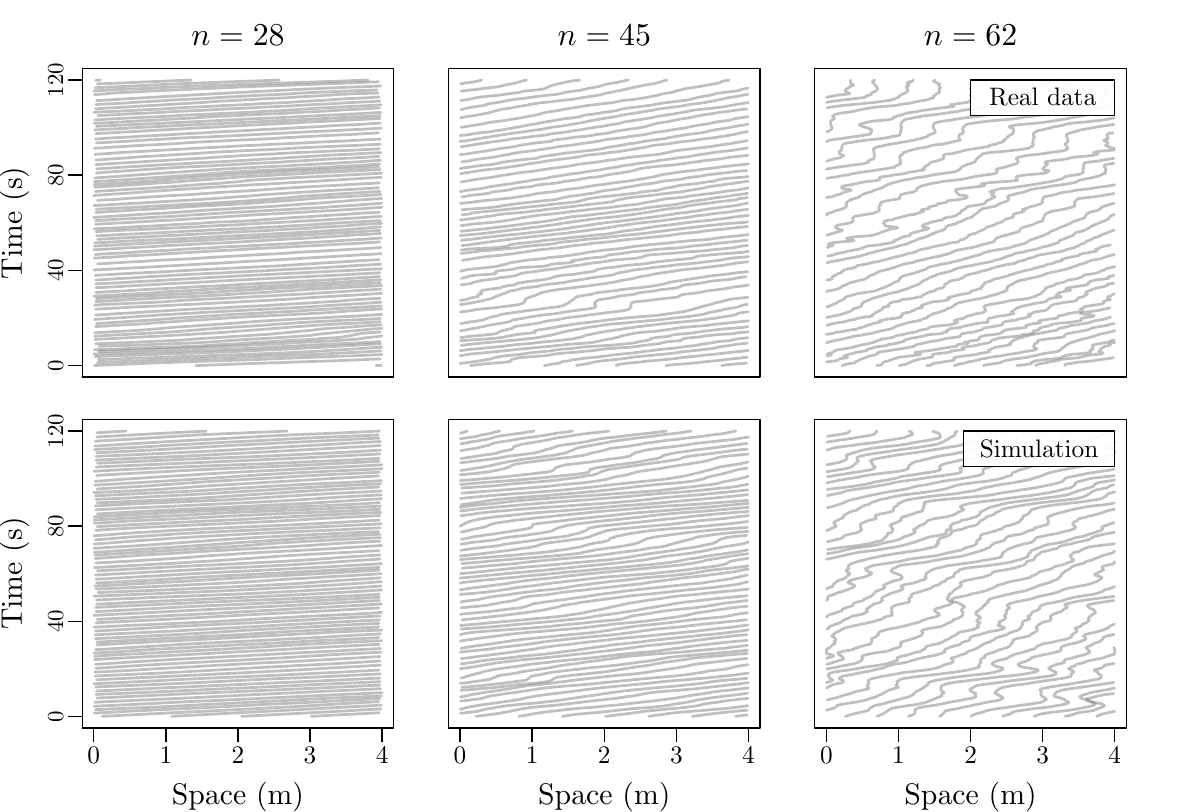}
\caption{Real (top panels) and simulated (bottom panels) trajectories
  for different densities.  The initial configuration is
  homogeneous.}
\label{fig:5}       
\end{figure}

Fig.~\ref{fig:5} presents real trajectories for experiments with $28$,
$45$ and $62$ participants (see \cite{Tordeux2016}) and simulations
with the stochastic model.  
The simulations are in good
agreement with the data.  Stop-and-go waves appear for semi-congested
($n=45$) and congested ($n=62$) states, while free states ($n=28$)
seem homogeneous in both empirical data and simulation. 

\section{Discussion}
\label{sec:4}

\begin{figure}[!ht]
\includegraphics{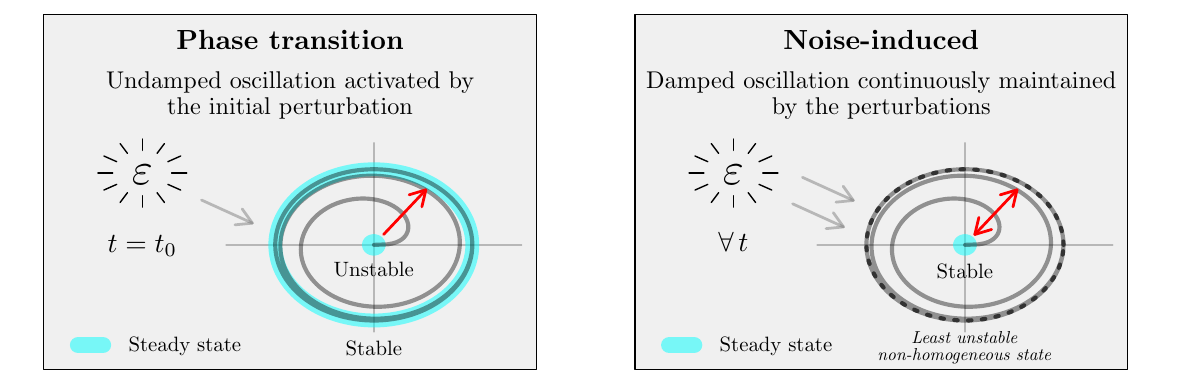}
\caption{Illustrative scheme for the modelling of stop-and-go dynamics
  with phase transition in the periodic solution (left panel) and the
  noise-induced oscillating behaviour (right panel).}
\label{fig:7}       
\end{figure}

We have proposed an original explanation for stop-and-go phenomena in
pedestrian flows as the consequence of a coloured noise in the
dynamics of the speed.  In this stochastic approach, oscillations in
the system 
are due to the perturbations provided by the noise.  Such a mechanism
qualitatively describes stop-and-go waves, especially when the system
is poorly damped.  The approach differs from classical deterministic
traffic models with inertia for which stop-and-go occurs due to
metastability and phase transitions to periodical dynamics (see
Fig.~\ref{fig:7}).

Two mechanisms based on relaxation processes are identified for the
description of stop-and-go waves.  In the novel stochastic approach,
the relaxation time is related to the noise and is estimated to
approximately 5~s \cite{Tordeux2016}.  The parameter corresponds to
the mean time period of the stochastic deviations from the
phenomenological equilibrium state.  Such a time can be high,
especially when the deviations are small and the spacings are high.
In the classical inertial approaches, the relaxation time is
interpreted as the driver/pedestrian reaction time and is estimated by
around 0.5 to 1~s.  Technically, such a parameter can not exceed the
physical time gap between the agents (around 1 to 2~s) without
generating unrealistic (colliding) behaviour and has to be set
carefully.

\begin{acknowledgement}
The authors thank Prof.\ Michel Roussignol for his help in the formulation of the model. 
Financial support by the German Science Foundation under grant SCHA 636/9-1 is gratefully acknowledged.
\end{acknowledgement}

\end{document}